\documentclass[prl,twocolumn,superscriptaddress,showpacs]{revtex4}
\usepackage{graphicx,bm,amssymb}
\newcommand{\pdag}{{\phantom{\dagger}}}
\begin{document}
\author{M. Pustilnik} 
\affiliation{School of Physics, Georgia Institute of Technology, Atlanta, GA 30332}
\author{L.I. Glazman} 
\affiliation{William I. Fine Theoretical Physics Institute, University of Minnesota, Minneapolis, MN 55455}
\author{W. Hofstetter}
\affiliation{Lyman Laboratory, Harvard University, Cambridge, MA 02138}
\title{Singlet-triplet transition in a lateral quantum dot}

%\date{\bf\today}

\begin{abstract}
  We study transport through a lateral quantum dot in the vicinity of
  the singlet-triplet transition in its ground state. This transition,
  being sharp in an isolated dot, is broadened to a crossover by the
  exchange interaction of the dot electrons with the conduction
  electrons in the leads. For a generic set of system's parameters,
  the linear conductance has a maximum in the crossover region. At
  zero temperature and magnetic field, the maximum is the strongest.
  It becomes less pronounced at finite Zeeman splitting, which leads
  to an increase of the background conductance and a decrease of
  the conductance in the maximum.
\end{abstract}


\maketitle

The Kondo effect in transport through quantum dots manifests itself in a
dramatic increase of the linear conductance at temperatures below a
certain characteristic scale $T_K$ (the Kondo temperature).  In the
simplest case~\cite{kondo}, a quantum dot behaves essentially as a
magnetic impurity with spin 1/2 coupled via exchange interaction to
two conducting leads~\cite{GP_review}. However, the energy scale for
intradot excitations is much smaller than the corresponding scale for
real magnetic impurities. Moreover, the tunability of this scale in
quantum dot devices allows one to explore various flavors of the Kondo
effect inaccessible with usual magnetic impurities~\cite{induced_review}. 

A transition between singlet and triplet states in an almost isolated dot 
was demonstrated~\cite{Tarucha} in a ``vertical'' device. At stronger
dot-lead tunneling, the conductance across the dot has a pronounced
maximum at the singlet-triplet transition~\cite{Sasaki}. The maximum
can be interpreted~\cite{induced_review,EN,ST} as the Kondo effect
with the Kondo temperature enhanced in the vicinity of the transition.
In this paper, we investigate the conductance in a ``lateral'' quantum
dot device in the vicinity of a singlet-triplet transition. The presented 
below theory of the Kondo effect in such devices may help in the 
interpretation of the recent experiments~\cite{vanderWiel,Kogan}.  

The main difference between vertical dots and lateral ones is that in 
the latter case the number of electronic modes coupled to a quantum 
dot is well defined. A lateral quantum dot is formed by gate depletion 
of a two-dimensional electron gas at the interface between two 
semiconductors. In this geometry the dot-leads junctions act as 
electronic waveguides. Potentials on the gates control the waveguide 
width, and, therefore, the number of propagating electronic modes the 
waveguides support: by making the waveguide narrower one 
pinches the propagating modes off. When the very last propagating 
mode nears its pinch-off, the system enters the Coulomb blockade 
regime. Accordingly, in this regime each dot-lead junction supports 
a \textit{single} electronic mode~\cite{MF}. 

Typically, the charging energy of the dot $E_C$ is large compared to 
the mean single-particle level spacing in it, $\delta E$, which in turn 
guarantees~\cite{GP_review} that $\delta E$ is large compared to $T_K$: 
$T_K \ll \delta E \ll E_C$. This separation of the energy scales 
allows us to simplify the problem further by assuming that conductances of 
the dot-lead junctions are small. The simplification will not affect the properties 
of the system in the Kondo regime as long as $T_K$ remains the smallest 
energy scale in the problem. On the other hand, the coupling between the dot 
and the leads can now be described within the tunneling Hamiltonian framework. 
The microscopic Hamiltonian of the system can then be written as a sum of three 
distinct terms,
\begin{equation} 
H = H_{\rm leads} + H_{\rm dot} + H_{\rm tunneling},
\label{model}
\end{equation} 
which describe, respectively, free electrons in the leads, isolated quantum 
dot, and the tunneling between the dot and the leads. With only one electronic 
mode for each dot-lead junction taken into account, %(see the discussion above), 
the first and the third terms in the r.h.s. of Eq.~(\ref{model}) become
\begin{eqnarray} 
H_{\rm leads} &=& 
\sum_{\alpha k s}\xi^\pdag_k c^\dagger_{\alpha k s} c^\pdag_{\alpha k s}, 
\quad \alpha=R,L
\label{H_leads} \\
H_{\rm tunneling} &=& \sum_{\alpha n k s} t^\pdag_{\alpha n} 
c^\dagger_{\alpha k s} d^\pdag_{ns} 
+ {\rm H.c.}
\label{H_tunneling}
\end{eqnarray} 
A generic model of an isolated quantum dot [the second term in the r.h.s. 
of Eq.~(\ref{model})] can be written as~\cite{ABG}
\begin{equation} 
H_{\rm dot} = \sum_{ns} \epsilon_n^\pdag d^\dagger_{ns} d^\pdag_{ns}
+ E_C \left(\hat N - \mathcal{N}\right)^2 - E_S\hat{\bm S}^2 - B\hat S^z, 
\label{H_dot}
\end{equation} 
where 
$\hat N =\sum_{ns} d^\dagger_{ns} d^\pdag_{ns}$
and
$\hat{\bm S} 
= \frac{1}{2}\sum_{nss'} 
d^\dagger_{ns} 
\hat{\bm{\sigma}}_{ss'}
d^\pdag_{ns'}$ 
are operators of the total number of electrons on the dot, and of the 
dot's spin, respectively. 
The parameter $\mathcal{N}$ in Eq.~(\ref{H_dot}) is proportional to 
the potential on the capacitively coupled gate electrode and 
controls the number of electrons, $N = \langle\hat N\rangle$, on the dot. 
We will assume that $\mathcal{N}$ is tuned to a Coulomb blockade
valley with an even number of electrons $N$. The third term in 
Eq.~(\ref{H_dot}) describes exchange interaction within the dot.  
Finally, the last term in Eq.~(\ref{H_dot}) represents the Zeeman effect 
of an external magnetic field, with $B$ being the Zeeman energy. 

We consider a dot in a state far from the ferromagnetic
instability~\cite{ABG}: the exchange energy $E_S$ is small compared to
the mean level spacing $\delta E$. Under this condition the ground
state of an isolated dot with even $N$ is almost always a singlet. The
only exception occurs when the level spacing
$\varepsilon=\epsilon_{+1}-\epsilon_{-1}$ between the highest occupied
($n=-1$) and the lowest empty ($n=+1$) single-particle energy levels
is anomalously small, $\varepsilon\sim E_S \ll \delta E$.  The gain in
the exchange energy associated with a formation of the triplet state
may then be sufficient to overcome the loss in the kinetic energy (cf.
the Hund's rule in atomic physics).  The triplet state is formed
through a redistribution of two electrons between the levels $n=\pm
1$. Since for an isolated dot the occupation $\sum_s d^\dagger_{ns}
d^\pdag_{ns}$ on each single-particle energy level is a constant of
motion, the redistribution must involve tunneling between the levels
$n=\pm 1$ and the leads. At low energies ($B,T\ll \delta E$) tunneling
to all other ($n\neq \pm 1$) energy levels can be neglected in the
vicinity of the singlet-triplet transition ($\varepsilon\sim E_S$).
Accordingly, in this regime the Hamiltonian of the dot Eq.~(\ref{H_dot})
can then be further truncated to that of a two-level system with
$\mathcal{N} = 2$ and
\begin{equation}
\epsilon_n = n\varepsilon/2, \quad n=\pm 1.
\label{TLS}
\end{equation}

In quantum dot systems based on GaAs the value of $\varepsilon$ can 
be controlled by a magnetic field $H_{\perp}$ applied perpendicular 
to the plane of the dot~\cite{Sasaki,vanderWiel}. Because of the smallness 
of the electron effective mass, even a weak field $H_{\perp}$ 
has a very strong orbital effect. At the same time, smallness of g-factor 
in GaAs ensures that the corresponding Zeeman splitting remains 
small~\cite{induced_review}. By linearizing $\varepsilon(H_\perp)$ in 
the vicinity of the transition, one can make a direct comparison of the 
experimental data with the (calculated) linear conductance across the system 
$G = (2e^2/h)g(\varepsilon)$. In addition, one can apply a strong in-plane
field $H_{\parallel}$, which would allow the study of  the conductance 
dependence on Zeeman energy $B$. The temperature ($T$) and field 
($B$) dependences of the linear conductance, and the bias dependence 
of the differential conductance are qualitatively similar to each other. However, 
the dependence $g(\varepsilon, B)$ at $T=0$ is easier to calculate, and it is 
this function we address here.

Tunneling, see Eq.~(\ref{H_tunneling}), couples the two-level system, 
Eqs.~(\ref{H_dot}) and (\ref{TLS}), to the two leads. The four tunneling 
amplitudes $t_{\alpha, \pm 1}$ form a $2\times 2$ matrix
\[
\hat t = \left(
\begin{array}{cc}
t_{L,+1} &  t_{R,+1} \\
t_{L,-1}  &  t_{R,-1} 
\end{array}
\right).
\]
In the {\it special case} when one of the eigenvalues of $\hat t$ is zero, 
while another one is finite~\cite{HS}, the dot effectively  interacts only 
with a {\it single} species (a single ``channel'') of conduction electrons. 
A single channel can screen only half of the dot's spin when it is in the 
triplet state~\cite{NB}. Accordingly, the system should exhibit a quantum 
phase transition~\cite{VBH}: the ground state changes its symmetry from 
a singlet to a doublet as $\varepsilon$ decreases below a certain value, 
$\varepsilon_C\sim E_S$. The conductance is then strongly 
$\varepsilon$-dependent~\cite{HS}: 
$g(\varepsilon,0) \propto \theta(\varepsilon_C -\varepsilon)$. 
At $\varepsilon < \varepsilon_C$ (when the dot is in the triplet state) 
the conductance is a monotonically decreasing function of $B$, while at 
$\varepsilon > \varepsilon_C$ the conductance first increases, and then 
drops with the increase of $B$~\cite{HS}. 

In the {\it general situation}, however, both eigenvalues of $\hat t$ are 
finite. Therefore, the dot is coupled to {\it two} electronic channels, 
which is sufficient to fully screen the it's spin~\cite{NB}. As the result, 
the ground state of the system is a {\it singlet} at all 
$\varepsilon$~\cite{ST,Izumida2,real}. In other words, when the dot 
is coupled to the leads, the singlet-triplet transition turns to a {\it crossover}. 
In order to study this crossover, we focus on a 
special subset~\cite{Izumida2,Izumida1} of the matrices $\hat t$ 
parametrized as
\begin{equation}
\hat t = \frac{1}{\sqrt 2}\left(
\begin{array}{cc}
v_{+1} &  v_{+1} \\
v_{-1}  &  - v_{-1} 
\end{array}
\right) .
\label{amplitudes}
\end{equation}
Obviously, for $v_{\pm 1}\neq 0$ both eigenvalues of $\hat t$ are 
finite, so the choice Eq.~(\ref{amplitudes}) captures the essential physics 
of the system. 

Since the ground state of the system is not degenerate, electrons scatter 
elastically at $T=0$. The amplitudes of scattering $\mathbb{S}_{s;\alpha\alpha'}$ 
of an electron with spin $s$ from lead $\alpha '$ to lead $\alpha$ form the 
scattering matrix $\hat\mathbb{S}_s$.  The $2\times 2$ unitary matrix 
$\hat\mathbb{S}_s$ can be diagonalized by a rotation in the $R-L$ space 
to the new basis of channels $n=\pm 1$,
\begin{equation}
U \hat\mathbb{S}_s U^\dagger = {\rm diag}\left\{e^{2i\delta_{ns}}\right\}, 
\quad
U = e^{i\vartheta_0 \tau^y}e^{i \varphi_0 \tau^z}. 
\label{diag}
\end{equation}
Here $\tau^i$ are the Pauli matrices acting in the $R-L$ space 
($\tau_+ = \tau_x + i\tau_y$ transforms $L$ to $R$). In general, 
the angles $\vartheta_0$ and $\varphi_0$ in Eq.~(\ref{diag}) depend 
on the parameters of the microscopic Hamiltonian, and, in particular, 
on the values of $\varepsilon$ and $B$. Here comes the key advantage 
of Eq.~(\ref{amplitudes}): with this choice of the tunneling amplitudes, 
the angles are parameter-independent constants: $\vartheta_0 = \pi/4$, 
$\varphi_0 =0$. Indeed, when written in terms of the operators 
\[
\psi_{nks} = \frac{1}{\sqrt 2}\left(n c_{Rks} + c_{Lks}\right),
\quad
n=\pm 1,
\]
[corresponding to $\vartheta_0 = \pi/4$ and $\varphi_0 =0$ in 
Eq.~(\ref{diag})], the Hamiltonian (\ref{model})-(\ref{amplitudes}) 
assumes the form
\begin{equation}
H = 
\sum_{n k s} \xi^\pdag_k \psi^\dagger_{nks} \psi^\pdag_{nks}
+ H_{\rm dot} 
 + \sum_{n k s}  v^\pdag_{n} \left(\!\psi^\dagger_{nks} d^\pdag_{ns} 
+ {\rm H.c.}\!\right)
\label{H_new}
\end{equation}
with $H_{\rm dot}$ given by Eqs.~(\ref{H_dot}), (\ref{TLS}). 
For each $n$ and $s$ this Hamiltonian commutes with the operator 
\[
\hat \mathbb{N}_{ns} 
= \sum_{k} \psi^\dagger_{nks} \psi^\pdag_{nks} + d^\dagger_{ns} d^\pdag_{ns}
\]
of the total number of electrons with spin $s$ on the ``orbital" $n$. Since 
$\hat \mathbb{N}_{ns}$ commute with each other, the (non-degenerate) 
ground state of $H$ is also an eigenstate of $\hat \mathbb{N}_{ns}$. This 
implies that single-particle correlation functions are diagonal in $n$ and $s$, 
e.g., $\langle\psi^\pdag_{nks}(t)\psi^\dagger_{n'k's'}(0)\rangle 
\propto \delta_{nn'}\delta_{ss'}$. Therefore the scattering matrix, which 
can be expressed via retarded single-particle correlation functions, is 
diagonal in $n$ and $s$ as well.

The dimensionless conductance $g$ at $T=0$ is related to the off-diagonal 
elements of the scattering matrix by the Landauer formula
$g = \frac{1}{2}\sum_s \left|\mathbb{S}_{s;RL}^2\right|$.  
With the help of Eq.~(\ref{diag}) the conductance can be expressed via 
the scattering phase shifts $\delta_{ns}$ at the Fermi energy:
\begin{equation}
g = \frac{1}{2}\sum_s\sin^2\Delta_s, 
\quad
\Delta_s = \delta_{+1,s} - \delta_{-1,s}.
\label{conductance}
\end{equation}
(here we took into account that $\vartheta_0=0$).  The scattering
phase shifts $\delta_{ns}$ in Eqs.~(\ref{diag}) and (\ref{conductance})
are obviously defined mod $\pi$ (that is, $\delta_{ns}$ and
$\delta_{ns} +\pi$ are equivalent). This ambuguity can be removed by
setting the values of the phase shifts corresponding to $\mathcal{N} =
-\infty$ (when the dot is empty) to $0$. With this convention, the
phase shifts at a finite $\mathcal{N}$ are expressed 
via the Friedel sum rule in terms of the ground state occupation numbers 
$N_{ns} = \langle d^\dagger_{ns}d^\pdag_{ns}\rangle$,
\begin{equation}
\delta_{ns} = \pi N_{ns}. 
\label{Friedel}
\end{equation}
Equation (\ref{Friedel}) is exact in the limit of infinite bandwidth. 
Alternatively, the phase shifts can be extracted directly from the 
finite-size spectra obtained by the numerical renormalization group~\cite{HZ}.

In the Coulomb blockade valley the number of electrons 
on the dot is fixed: $N = \sum_{ns} N_{ns} = 2$. Accordingly, 
the phase shifts satisfy
\begin{equation}
\sum_{ns} \delta_{ns} = 2\pi.
\label{valley}
\end{equation}
Additional relation for the phase shifts follows from the  invariance of 
the Hamiltonian (\ref{model})-(\ref{H_dot}) with respect to the 
transformation $(s,B)\to (-s,-B)$:
\begin{equation}
\delta_{ns}(B) = \delta_{n,-s}(-B).
\label{symmetry}
\end{equation}

\begin{figure}[h]
\includegraphics[width=0.7\columnwidth]{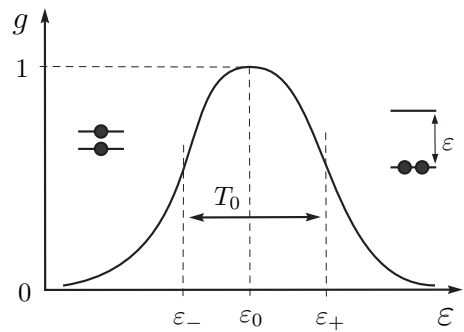}
\caption{Dimensionless conductance across the dot at $T=0$ and $B=0$.
The conductance decreases with $B$ at $\varepsilon_- < \varepsilon < \varepsilon_+$
and increases with $B$ outside this interval.
\label{zero_field}
}
\end{figure}

Consider first the limit of zero field, $B=0$.
From Eqs.~(\ref{valley}) and (\ref{symmetry}) it follows that
\begin{equation}
\delta_{ns}(0) = \delta_{n,-s}(0) = \delta_n,
\quad
\sum_n\delta_n = \pi.
\label{delta_0}
\end{equation}
At the triplet side of the crossover $\varepsilon\ll E_S$ both levels
in the dot are singly occupied, so that Eq.~(\ref{Friedel}) yields
$\delta_{n} = \pi/2$.  On the contrary, at $\varepsilon\gg E_S$ the
level $n=-1$ is doubly occupied, while the level $n=+1$ is empty,
which corresponds to $\delta_{-1} = \pi$, $\delta_{+1} = 0$. When
$\varepsilon$ is tuned through the crossover, the difference of the
phase shifts, $\Delta_s=\delta_{+1} - \delta_{-1}$ [see
Eq.~(\ref{conductance})], monotonically decreases from $0$ to $-\pi$.
It follows then from Eq.~(\ref{conductance}) that the dependence of
the dimensionless zero-field conductance $g(\varepsilon,0)$ on
$\varepsilon$ is nonmonotonic. The conductance reaches its maximum
$g=1$ at some $\varepsilon = \varepsilon_0$ corresponding to
$\Delta_s(\varepsilon_0) = -\pi/2$, and falls off monotonically with
the distance $|\varepsilon-\varepsilon_0|$ to this point, see
Fig.~\ref{zero_field}. The energy $\varepsilon_0\simeq 2E_S$ may be
identified with the center of the crossover region. The conductance
$g\sim 1$ within this region. Later on we will relate the width of the 
crossover region $T_0$ to the parameters of the 
Hamiltonian~(\ref{H_dot}),(\ref{H_new}).

The effect of the Zeeman energy $B$ on the conductance accross the
dot depends on how far its parameters are from the crossover point. 
In order to study the influence of a small $B$, one can expand 
the phase shifts in a series~\cite{Nozieres}. Taking into account 
Eqs.~(\ref{valley}), (\ref{symmetry}), and (\ref{delta_0}), we obtain
\begin{equation}
\delta_{ns}(\varepsilon,B) = \delta_{n} 
+ s\left({B}/{T_n}\right) 
+ n \left({B}/{T'}\right)^2 + O(B^3),
\label{scales} 
\end{equation}
where $n=\pm 1$, $s=\pm 1$, and $\delta_n$, $T_n$, and $T'$ depend on 
$\varepsilon$. The sign of the linear in $B$ term in Eq.~(\ref{scales}) 
is fixed by the following argument. The spin polarization $N_{n,+1}-N_{n,-1}$ 
of electrons on each level ($n=\pm 1$) grows proportionally to $B$. It follows 
then from Eqs.~(\ref{Friedel}) and (\ref{scales}) that $T_n>0$ for $n=\pm 1$. 
In the next order in $B$, magnetic field favors a triplet over the singlet state 
of the dot. Therefore, the difference $N_{+1,s} -N_{-1,s}$ increases with 
$|B|$, which fixes the sign of the second order in $B$ contribution in 
Eq.~(\ref{scales}). 

%%%%%%%%%%%%%%%%%%%%%%%%%%%%%%%%
%%%%%%%%%%%%%%%%%%%%%%%%%%%%%%%%

Substitution of Eq.~(\ref{scales}) into Eq.~(\ref{conductance}) yields 
the low-$B$ asymptotics of the conductance,
\begin{equation}
g(\varepsilon,B) = g + (1-2g) \left(B/B_0\right)^2 + 4\sqrt{g(1-g)}\left(B/T'\right)^2, 
\label{small_B}
\end{equation}
with $g = g(\varepsilon,0)$ and
\begin{equation}
\frac{1}{\,B_0(\varepsilon)} = \left|
\sum_{n=\pm 1}\frac{n}{T_n(\varepsilon)}\right|\,.
\label{gap} 
\end{equation}
Around the crossover point $g\approx 1$, the last term in 
Eq.~(\ref{small_B}) vanishes, and
\begin{equation}
g(\varepsilon,B) \approx 
1 - \left(B/B_0\right)^2, 
\quad
|\varepsilon-\varepsilon_0|\ll T_0.
\label{at_crossover}
\end{equation}
Away from the crossover ($g\ll 1$), Eq.~(\ref{small_B}) yields
\begin{equation}
g(\varepsilon,B) \approx 
g(\varepsilon,0) + \left(B/B_0\right)^2, 
\quad
|\varepsilon-\varepsilon_0|\gg T_0\,.
\label{away_crossover}
\end{equation}

It is clear from Eqs.~(\ref{at_crossover}) and (\ref{away_crossover})
that the conductance varies with the magnetic field in the opposite
directions in the vicinity and far away from the crossover point. 
It should be emphasized that this is a generic property rather than a consequence 
of the specific choice of the model, Eq.~(\ref{amplitudes}).
However, the precise borders of the crossover region $\varepsilon = \varepsilon_\pm$, 
see Fig.~\ref{zero_field}, as well as zero-field conductances $g(\varepsilon_\pm, 0)$ 
at these points are model dependent.  Note also that Eq.~(\ref{small_B}) is 
applicable only at $B\ll B_0(\varepsilon)$. The conductance at higher fields 
is a nonmonotonic function of $B$, see Refs.~\cite{real,Izumida1,HZ,Zeeman}.

The Hamiltonian Eq.~(\ref{H_new}) is identical to that employed
previously~\cite{ST} to study transport through a {\it vertical} dot.
Accordingly, the thermodynamic properties of a lateral dot coincide
with those of a vertical dot. However, the existing experiments test
transport properties rather than thermodynamics. The electron current
operators are different for the two models, hence the dependences
$g(\varepsilon,B)$ are different as well. For instance, the
conductance through a vertical dot at $T=0$ is large ($\sim 4e^2/h$)
at the triplet side of the crossover and decreases with $B$, in
contrast with Eq.~(\ref{away_crossover}).

At $|\varepsilon-\varepsilon_0|\gg T_0$ the parameters $g(\varepsilon,0)$ 
and $B_0(\varepsilon)$ entering Eq.~(\ref{away_crossover}) can be 
estimated with the help of the perturbative renormalization group~\cite{ST}.
In this regime the $\varepsilon$-dependence of all observable quantities 
is governed by the parameter
\begin{equation}
x(\varepsilon) = (\varepsilon - \varepsilon_0)/T_0,
\quad
1\ll \,|x(\varepsilon)|  \ll \delta E/T_0 .
\label{X} 
\end{equation}
The estimated width of the crossover region satisfies~\cite{ST} 
\begin{equation}
\ln(\delta E/T_0) \approx {\frac{E_C C(\gamma)}{\nu(v_{+1}^2 + v_{-1}^2)}},
\quad
\gamma = |v_{+1}/v_{-1}|,
\label{T_0} 
\end{equation}
where $\nu$ is density of states in the leads.
The function $C(\gamma)$ has a minimum $C\approx 0.36$ at $\gamma =1$ 
and goes over to $1$ as $\gamma\to 0,\infty$. The scale $T_0$ also plays the part 
of the Kondo temperature as it determines the $T$-dependence of the 
conductance in the crossover region: the peak in $g(\varepsilon)$ 
disappears at $T\gg T_0$. Typically, $T_0$ is larger than the Kondo temperatures 
in the nearby Coulomb blockade valleys with odd number of electrons 
on the dot~\cite{EN,ST}. 

The zero-$B$ conductance
entering Eq.~(\ref{small_B}) away from the crossover point (at $|x|\gg 1$) is
\begin{equation}
g(\varepsilon,0) \propto (\ln |x|)^{-2} .
\label{g0} 
\end{equation}
Finally, the characteristic magnetic field $B_0 (\varepsilon)$ defined in
Eq.~(\ref{gap}) has different asymptotes at the triplet and
singlet sides of the crossover:
\begin{eqnarray}
{B_0}/{T_0} \propto \left\{
\begin{array}{lr}
|x|^{-\lambda} (\ln |x|)^{2\lambda/(\lambda + 1)}, 
& x<0,
\\ \\
x(\ln x)^{2(\lambda+2)/(\lambda + 1)},
& x > 0
\end{array}
\right.
\label{gaps}
\end{eqnarray}
with $\lambda = 2 +\sqrt{5}\approx 4.2$ and $|x|\gg 1$.

The dependence of the differential conductance $dI/dV$ on the
source-drain bias $V$ at $eV\gg T,B$ is qualitatively similar to the
dependence of the linear conductance on $B$ at $B\gg T$.  The
similarity stems from the fact that $dI/dV$ is determined by the
electron transmission coefficient at energies $\omega\sim eV$, where
$\omega$ is measured from the Fermi level~\cite{GP_review}.  The
dependences of the transmission coefficient on $\omega$ and $B$ are
controlled by a single energy scale; the $\omega=0$ limit of the
transmission coefficient governs~\cite{GP_review,real} the linear
conductance $g(\varepsilon,B)$.  Therefore, Eq.~(\ref{at_crossover})
implies also that $dI/dV$ decreases with $V$ in the crossover region
shown in Fig.~\ref{zero_field}. On the contrary, away from the
crossover the conductance is small, $dI/dV\sim
(2e^2/h)g(\varepsilon,0)$, in the domain $V\ll \Delta V\sim B_0/e$ and
is increasing with $V$, cf. Eq.~(\ref{away_crossover}).  The
``window'' $\Delta V$ is narrow at the triplet side of the crossover
but broadens approximately linearly with the distance to the crossover
at the singlet side of it, see Eq.~(\ref{gaps}). This is in agreement with numerical 
simulations~\cite{HZ}. 

The zero-bias suppression of $dI/dV$ was observed in recent
experiments~\cite{vanderWiel,Kogan}.  A measurement in the limit of
very weak tunneling, necessary for the proper characterization of the
spin state of the dot, was not possible for the studied devices.
However, the observed~\cite{vanderWiel} asymmetric behavior of $\Delta
V$ across the point where the conductance has a maximum is precisely
the expected behavior in the vicinity of the singlet-triplet
crossover.  The asymmetric behavior of $\Delta V$ across the crossover
may explain the observed~\cite{Kogan} splitting of the Kondo peak in a
certain range of gate voltages.

%\begin{acknowledgments}

We benefited from discussions and correspondence 
with S. De Franceschi, M.A. Kastner, A. Kogan, L.P. Kouwenhoven, 
H. Schoeller, W.G. van der Wiel, and G. Zarand. 
Two of us (M.P. and L.I.G.) are grateful to the Kavli Institute for Theoretical 
Physics at USCB and to the Institute for Nuclear Theory at the University 
of Washington for their hospitality.  
This work was supported by NSF grants DMR97-31756, DMR02-37296, and
EIA02-10736. W.H. acknowledges financial support from the German
Science Foundation (DFG), and L.I.G thanks the Institute for
Strongly Correlated and Complex Systems at BNL for hospitality and
support.

%\end{acknowledgments}

\end{document}